\documentstyle[aps,preprint]{revtex}

\newcommand{\Ref}[1]{Ref.~\cite{#1}}

\newcommand{\be}{\begin{equation}}
\newcommand{\ee}{\end{equation}}
\newcommand{\ba}{\begin{eqnarray}}
\newcommand{\ea}{\end{eqnarray}}
\newcommand{\nn}{\nonumber}
\newcommand{\Eq}[1]{Eq.~(\ref{#1})}

\newcommand{\vev}[1]{\left\langle #1 \right\rangle}

\renewcommand{\Im}{{\rm Im}}
\renewcommand{\Re}{{\rm Re}}

\newcommand{\prt}{\partial}

\renewcommand{\v}[1]{{\bbox #1}}

\newcommand{\up}{\uparrow}
\newcommand{\down}{\downarrow}

\newcommand{\al}{\alpha}
\newcommand{\bt}{\beta}
\newcommand{\del}{\delta}
\newcommand{\Del}{\Delta}
\newcommand{\eps}{\epsilon}

\newcommand{\la}{\lambda}

\newcommand{\om}{\omega}

\newcommand{\th}{\theta}

\newcommand{\si}{\sigma}

\def\journal #1, #2, #3, 1#4#5#6{{\sl #1~}{\bf #2}, #3 (1#4#5#6) }

\begin{document}
\widetext

\title{
Preformed Pairs,
$SU(2)$ Slave-boson Theory, and High $T_c$ Superconductors
}

\author{Ching-Long Wu$^{a,b}$, Chung-Yu Mou$^{a,b}$, Xiao-Gang
Wen$^{a,c}$,
Darwin Chang$^{a,b}$}
\address{
a) Physics Division, NCTS, P.O.Box 2-131, Hsinchu, 30043 Taiwan, ROC\\
b) Physics Department, National Tsing Hua University, Hsinchu, 30043,
Taiwan, ROC\\
c) Department of Physics, Massachusetts Institute of Technology,
Cambridge, MA 02139, USA
}

\maketitle

\widetext
\begin{abstract}
\rightskip 54.8pt A preformed-pair model was considered. The
quantum disordered phase of the $d$-wave superconducting state was
obtained by turning on an on-site repulsion. The low energy
effective theory of the quantum disordered phase was derived,
which was found to be qualitatively equivalent to the low energy
effective theory of the $SU(2)$ slave-boson theory in the
staggered-flux phase. Many physical properties of the disordered
phase at low energies were obtained, such as the Mott insulator
property at zero doping, spin gap and low superfluid density at
low doping.
\end{abstract}

\pacs{ PACS numbers:  74.25.Jb,79.60.-i,71.27.+a}


\narrowtext

The cuprate superconductors not only contain a high $T_c$ superconducting (SC)
phase, they also contain, in the neighborhood of the SC phase,
an antiferromagnetic (AF) phase (at and near zero doping), a spin-gap (SG)
phase (in underdoped regime), a non-Fermi liquid (NF) phase with a large Fermi
surface (near optimal doping), and a Fermi liquid (FL) phase (in overdoped
regime). (Of course the last three phases are not really different phases.
They can continuously cross over into each other without any phase
transition.) There are several important energy scales showing up in the above
phases. The most important scale is the charge gap $U\sim 2$eV at
zero doping. Such a charge gap, as the energy cost for the doubly occupied
site, also plays extremely important role in the underdoped regime.
The next energy scale is $J\sim 0.1$eV -- the interaction energy of spins.
Since $J\ll U$, the undoped cuprate is a Mott insulator. There is also
a spin gap appearing
at energy scale $\Del\sim 0.03$eV in the SG phase. The energy scale
associated with the superconductivity is given by
$T_c\sim 0.003 - 0.01$eV, whose value depends on the doping concentrations.
A theoretical model for the high $T_c$ superconductors (HTS) should address
the above
energy scales, or at the very least address the large charge gap $U$.

The slave-boson approach (both the $U(1)$ and $SU(2)$ theories)\cite{RVB}
to the HTS takes the large charge gap as the main input.
The model imposes a no-double-occupancy constraint, which
effectively sets the charge gap $U\to \infty$. This approach leads
to an effective theory which contains spinons (spin 1/2 neutral
fermions), holon (spinless charge $e$ bosons), and a gauge field.
The spinon, describing the spin degree of freedom, has an energy
scale $J$, the spin-gap phase can be explained by the
staggered-flux (sF) phase of spinons, which has an energy scale of
a fraction of $J$\cite{RVB,WL}.

After the photoemission experiments \cite{photoE}
which confirmed the spin gap and measured their momentum
dependence for underdoped HTS's, many groups \cite{ppair} proposed
a preformed-pair picture to explain the spin gap. Although the preformed-pair
picture is very natural in explaining the spin gap and
its momentum dependence at energy scale $\sim 25$meV, it is not clear how to
recover the large charge gap near 2eV within this picture,
{\it i.e.} it is not clear
how to explain the insulating properties at zero doping within the
preformed-pair picture.

Historically, a ``preformed-pair picture'' was first suggested by Anderson
and his group under the name of Resonating Valence Bound (RVB) state
\cite{RVB}.
The main difference between the RVB state and the above preformed-pair picture,
is that the RVB state also imposes the no-double-occupancy constraint.
Thus the RVB state is naturally a Mott insulator at half filling,

In this paper, we start from a preformed-pair model with $d$-wave pairing
(which addresses the spin
gap first), and study the quantum disordered $d$-wave (QDdW) phase
through a mathematical trick of
promoting the effective XY model (or the $O(2)$ non-linear
$\si$-model) to an anisotropic $O(3)$ non-linear $\si$-model.
It is difficult to study the disordered phase of XY model directly, because
the vortices, which are important in the disordered phase, are singular
objects in the XY model. However, as we will see, regarding the XY model as an
anisotropic $O(3)$ non-linear $\si$-model can overcome this difficulty.
We find that the  QDdW phase is described by
an effective theory with spin 1/2 neutral fermions, spinless charge $e$
bosons and a $U(1)$ gauge field. The effective theory gives rise to
an insulating phase at half filling with a finite charge gap. The optical
conductivity $\si(\om)=0$ for $\om$ below the charge gap if there is a
particle-hole symmetry.
Although our calculation is reliable only when the charge gap $U$ is less
than the spin gap $\Del$, the effective theory is well behaved and reasonable
even when we push the charge gap well beyond the spin gap. In this limit
our effective theory is very similar to effective theories of
sF phase obtained from the $SU(2)$ slave-boson approach \cite{WL}.
The close relation
between the QDdW phase
and the sF phase of the $SU(2)$ slave-boson theory
is not surprising. The underlying physics of the  sF phase in the
slave-boson theory is the RVB picture (note that the sF phase is gauge
equivalent to the $d$-wave paired phase of spinons within the $SU(2)$ slave
boson theory\cite{WL}).
The RVB state is nothing but a liquid of non-overlapping Cooper pairs.
It is only natural to see that the liquid of non-overlapping Cooper pairs
is also the QDdW phase.

A recent work \cite{BFN} also started from a preformed-pair model
and studied the QDdW phase (which was named the ``nodal phase'')
through a mathematical trick of duality transformation of
XY model. However, due to some unknown reasons, the effective theory
obtained in \Ref{BFN} for the nodal phase
has some physical results which are different
from the ones obtained from our effective theory. The most notable
difference is that the neutral spin 1/2 fermions do not carry the charge of
the gauge field, while in our approach they do carry the charge.

We start with a generalized Hubbard model which contains an on-site repulsion
$U$ and a nearest neighbor spin coupling:
\be
H=\sum_{\vev{ij}}
(  c^\dag_{i} t_{ij} c_{j} + h.c. ) + \frac{U}{2} \sum_i (n_i -1)^2
+J \sum_{\vev{ij}} \v S_i \cdot \v S_j 
\label{HgH}
\ee
Note that the Hubbard model can generate the above spin coupling with
$J=4t^2/U$.
Here we treat $U$ and $J$ as independent variables.
We will limit our discussion at zero temperature.  Let us ignore the
on-site repulsion $U$
at the moment. The nearest neighbor spin coupling causes
an attraction between a spin-up and spin-down electrons. We assume
that such an attraction leads to a $d$-wave SC state
described by the following mean field theory
\be
H=\sum_{\vev{ij}}
(  c^\dag_{i} t_{ij} c_{j} +
c^\dag_{i\al} \Del_{ij} \eps_{\al\bt} c^\dag_{j\bt} + h.c. )
\label{dwHeff}
\ee
where $\Del_{ij}=\Del_{ji}$ is the $d$-wave pairing amplitude.
As we turn on the on-site repulsion $U$, the $d$-wave pairing order parameter
$\Del_{ij}$ will have stronger and stronger quantum fluctuations. If we
assume
the fluctuation is mainly the phase fluctuation at a length scale larger
than a few lattice spacing, then the quantum fluctuations may destroy the
SC phase without closing the $d$-wave spin gap. This picture may explain
the $d$-wave-like spin gap in the normal state of underdoped cuprates.
Notice that the on-site $U$ can play two roles a) it increases the
superconducting
phase fluctuation and destroy the SC state, and b) it opens up a charge gap
at zero doping (at least when $U$ is very large).
Thus it is natural to guess that, at zero doping, the QDdW phase caused by
finite on-site repulsion $U$ is also an insulator with finite charge gap.
In the following, we will derive an effective theory for the
QDdW phase and argue that this is indeed the case.

In addition to the $SU(2)$ spin symmetry,
the Hamiltonian \Eq{HgH} has another $SU(2)$ symmetry at half filling
(when the chemical potential gives $\vev{n_i}=1$), 
which we call the
pseudo-spin symmetry. The $SU(2)$ pseudo-spin symmetry can be made explicit by
introducing the doublet
\be
\la_i =
\pmatrix{\la_{i1} \cr \la_{i2} \cr} =
\pmatrix{c_{i\up} \cr (-)^i c^\dag_{i\down} \cr}
\ee
which carries pseudo-spin 1/2.
(This symmetry is explicitly broken 
if we introduce a $-\frac{1}{4} n_i n_j$ term in  \Eq{HgH}. 
We shall consider the effect of such term later.)
We would like to {\em stress} that although our theory has the
pseudo-spin
symmetry, it is not essential to our calculations
and results. 
{\em Most results obtained in this paper are still
valid even without the symmetry.}

In terms of $\la$, the $d$-wave effective Hamiltonian \Eq{dwHeff}
can be rewritten as
\be
H=\sum_{\vev{ij}} \left(  \la^\dag_i (t_{ij}-(-)^i (i\Im \Del_{ij}
\si_1 + i\Re \Del_{ij} \si_2 )) \la_j + h.c. \right)
\ee

The dynamics of the $O(2)$ superconducting order parameter
$(\Re \Del_{ij}, \Im \Del_{ij} )$ is described by a 2D XY model,
whose disordered phase
contains vortices. To avoid this difficulty, we promote the $O(2)$
order parameter
to an $O(3)$ order parameter $\Del^d_{ij}(n_{ij}^1, n_{ij}^2,
n_{ij}^3)$:
\ba
 H&=&\sum_{\vev{ij}} \left(  \la^\dag_i (t_{ij}-
i\Del^d_{ij} \v n_{ij} \cdot \v \si (-)^i) \la_j + h.c. \right)
+ \sum_{\vev{ij}} \frac{1}{2g} (\v n_{ij})^2.
\label{o3Heff}
\ea
Here $\Del^d_{ij}$ is real and is the gap function for a constant $d$-wave
order parameter, $\v n_{ij}$ is a unit vector.
We have included the quadratic $(\v n_{ij})^2$ term to implement the
constraint,
and the bare value of $1/g$ is $\Delta^2/2J$.
Note that in the presence of the pseudo-spin symmetry, the ground
state is described by the $O(3)$ order parameter $\v n_{ij} $
instead of the $O(2)$ superconducting order parameter.
$ \v n_{ij}$ is a pseudo-spin vector and \Eq{o3Heff}
has the same pseudo-spin rotation symmetry as in \Eq{HgH}.

$n_{ij}^{1,2}$ describe the phase fluctuations of the $d$-wave order parameter.
$n_{ij}^3$ describe a new kind of fluctuation. To have a better understanding
about the $n_{ij}^3$ fluctuations, let us rewrite \Eq{o3Heff} in terms of the
electron $c_i$ operators: \ba H&=&\sum_{\vev{ij}} \left( c^\dag_{i\alpha}
(t_{ij} - i(-)^i\Del^d_{ij} n_{ij}^3 ) c_{j\alpha} + h.c. \right) \nn\\ &&
+ \left(c^{\dagger}_{i\al}\Del^d_{ij}(n_{ij}^2+in_{ij}^1)
\eps_{\al\bt}c^{\dagger}_{j\bt}+h.c.
\right) + \sum_{\vev{ij}} \left(\frac{1}{2g} (\v n_{ij})^2
 \right)
\label{o3cHeff}
\ea
It is clear that the $n_{ij}^3$ fluctuations correspond to staggered flux
fluctuations. 


To understand the dynamics of $\v n_{ij}$, we also need to know its temporal
part which is generated from the coupling to the fermions.
Since $n_{ij}^{1,2}$ carry electric charge $\pm 2e$, the  temporal
part has the following form as required by the electromagnetic gauge
invariance and the pseudo-spin symmetry:
\be
L_b=\sum_{\vev{ij}} \frac{a^2}{16\chi}
[(\del_{ab}\prt_t +{2e} A_0 T_{ab})n^{b}_{ij}]
(\del_{ab'}\prt_t +{2e} A_0 T_{ab'})n^{b'}_{ij}
\label{dtn}
\ee
where
$T=\pmatrix{
 i\si_2 &0\cr
 0&0\cr
},$ $a$ is the lattice constant, and $\chi$ is the compressibility
identified from the coefficient of $(e A_0)^2$,
$e A_0$ being the chemical potential.  The compressibility can be obtained
by integrating out the electron degree of freedom and is given by the integral
\be
  \frac{a^2}{16\chi}=\frac{a^2}{2\pi}
  \int{d^2k \frac{\Delta_k^2}{(\epsilon_k^2+\Delta_k^2)^{\frac{3}{2}}}}
\label{comp}
\ee


The next task is to obtain the effective theory for the QDdW
phase of \Eq{o3Heff}.
One way to obtain the effective theory is to set
$\v n_i=0$ in \Eq{o3Heff} since $\vev{\v n_i}=0$ in the disordered
phase. However the effective theory obtained this way is not a good
description of the disordered phase in our mind. The disordered phase that we
considered has the following picture. The vector $\v n_i$ varies slowly in
space and time.
Thus locally $\v n_i$ can be treated as a constant vector
which gives rise to a finite spin gap in fermionic quasi-particles
spectrum even in
the disordered state.
Thus it is better to do an expansion in the power of $\prt \v n$ instead of in
the power of $\v n$. The new expansion can be achieved through a
``local rotation'', which makes the $\prt \v n$ dependence explicit.
This ``local rotation'' has been used to study the disordered phase of the
antiferromagnetic state. \cite{WdAF}

%
%

The local rotation can be realized in the following way.
We first parameterize $\v n_{ij}$ in \Eq{o3Heff} as ${\bf
n}_{ij}=\frac{1}{2}\left( {\bf n}_{i}+{\bf n} _{j}\right) $. (This
parameterization ignore some short wavelength fluctuations.) Note
that in long wavelength approximation, ${\bf n}_{i}$ and ${\bf
n}_{j}$ point to approximately the same direction as ${\bf
n}_{ij}$, since they are nearest neighbors, therefore the ${\bf
n}_{ij}$ thus constructed can still be treated as a vector of unit
length. Next we apply a local rotation at each site $i$ such that
${\bf n}_{i}^{\prime }s$ are all rotated to the $z$-direction.
This is done by introducing two new fields $z_i$ and $\psi_i$
\cite{WdAF}:
\begin{eqnarray}
\label{nz}
{\bf n}_{i} &=& z_{i}^{\dagger }{\v \sigma }z_{i},\ \ \
z_{i}^{\dagger }z_{i} =1 \nonumber \\
\lambda _{i} &=& U_{i}\psi _{i}
\end{eqnarray}
where $U_{i}=\pmatrix{
z_{i1} & -z_{i2}^*  \cr
z_{i2} & z_{i1}^* 
}
$
, and
${\bf n}_{i}\cdot U_{i}^{\dagger }{\v \sigma
}U_{i}=\sigma _{3}$.

Now we can rewrite $ {\bf n}_{ij} \cdot \lambda _{i}^{\dagger }{\v
\sigma }\lambda _{j}
=\frac{1}{2}\psi _{i}^{\dagger }\left\{ \sigma _{3},U_{i}^{\dagger
}U_{j}\right\} \psi _{j}
$.
Even when $\v n_i$ is very close to $\v n_j$, $z_{i}$ and $z_{j}$
can still differ by a finite $U(1)$ phase. Thus, to the lowest
order approximation, the main difference between $z_{i}$ and
$z_{j}$ is the $U(1)$ phase. Define $ a_{0i}=-iz^\dag_i \prt_t
z_i$ and $ z_{i}^{\dagger }z_{j}\simeq e^{ia_{ij}}$ , then
$U_{i}^{\dagger }U_{j}\simeq e^{ia_{ij}\sigma _{3}}$.
The fermion part of the effective Lagrangian becomes
\ba
L_f &=&\sum_i i\psi _{i}^{\dagger
}(\prt_t+ia_0\si_3) \psi _{i} \nn\\ && -\sum_{\vev{ij}}\left( \psi
_{i}^{\dagger } \left( t_{ij}-i\left( -\right)
^{i}\Delta^d_{ij}\sigma _{3}\right)e^{ia_{ij}\sigma _{3}} \psi
_{j}+c.c. \right) \label{cp1f}
\ea
Next we look at the boson part
\ba
L_{b}&=& -\sum_{\vev{ij}}
\frac{1}{2g}\left( {\bf n}_{ij}\right) ^{2} = -\sum_{\vev{ij}}
\frac{1}{g}\left( 1+{\bf n}_{i}\cdot {\bf n}_{j}\right),
\label{cp1b}
\ea
where ${\bf n}_{i}\cdot {\bf n}_{j}$ can be rewritten in terms of $z_i$ as
\ba
{\bf n}_{i}\cdot {\bf n}_{j}
&\simeq& 2\left( z_{i}^{\dagger }e^{-ia_{ij}}z_{j}+c.c.\right) -1
\ea
Although $L_b$ does not contain it, a temporal term for $z_i$
can be generated by integrate out the fermions. 
\Eq{cp1f} and \Eq{cp1b} define a $CP^1$ model (coupled to
fermions) on the lattice.
%
Using the standard approach to the disordered phase of $CP^1$ model\cite{CP1}
, we obtain the following effective Lagrangian (for the QDdW phase)
\ba
&&L_{eff} = \sum_i i\psi^\dag_i (\prt_t + ia_0 \si_3 ) \psi_i \nn\\
&&-\sum_{\vev{ij}} \left(  \psi^\dag_i \left(t_{ij}- i\Del^d_{ij} \si_3 (-)^i
\right) e^{ia_{ij}\si_3 } \psi_j  + h.c. \right) \nn\\ &+& \sum_i \frac{1}{g'}
\left(
|(\prt_t - ia_0 - i e A_0\si_3) z_i|^2 - \Del_c^2 z_i^\dagger z_i \right) \nn\\
&&-\frac{2}{g} \sum_{\vev{ij}} \left(  z^\dag_i
(e^{-ia_{ij}-i\frac{e}{c}A_{ij}\si_3}-\delta_{ij}) z_j + h.c. \right)  + ...
\label{Leff} \ea where $1/g'=a^2/2\chi$, and the $z_i^\dagger z_i$ term serves
as the Lagrange multiplier which frees the $z_i$ field from the $z^\dag_i
z_i=1$ constraint, it is also this term that gives the charge gap $\Delta_c$ in
our effective theory. The $|(\prt_t - ia_0 - i e A_0\si_3) z_i|^2 $ term comes
from the $(\prt_t \v n_{ij})^2$ term in \Eq{dtn}. The above low energy
effective Lagrangian for the QDdW phase is the main result of this paper.

{}From the definition of $z_i$ and $\psi_i$ in \Eq{nz}, we
note that $z_i$ carries pseudo-spin 1/2, and $\psi_i$ carries
pseudo-spin zero. For the spin and the charge, $z_i$
carries spin zero and charge $-e$ (or $e$), while $\psi_i$ carries
spin 1/2 (with $S_z=+1/2$) and charge zero. We see that $\psi_i$
and $z_i$ are very much like the spinon and the holon in the
$SU(2)$ slave-boson theory.

Also, our theory is originally expressed in terms of physical
variables $\v n_i$ and $\la_i$. Rewriting the theory in terms of
$\psi_i$ and $z_i$ introduces unphysical degree of freedom. The
theory is invariant under the local transformations
\be
z_i \to e^{-i\th_i(t)} z_i,\ \ \ \psi_i \to e^{i\th_i(t)\si_3 } \psi_i
\label{zpsith}
\ee
because $\v n_i$ and $\la_i$ are invariant under those transformations.
Thus it is not surprising to see that the effective theory $L_{eff}$ is a
$U(1)$ gauge theory, and is invariant under the gauge transformation
\Eq{zpsith} and
\be
a_{ij}\to a_{ij}+\th_i-\th_j,\ \ \ a_{0i} \to a_{0i}-\prt_t \th_i
\ee
It is also clear that the gauge field $a_\mu$ carries no spin,
charge and pseudo-spin quantum numbers.

We would like to discuss the above effective Lagrangian in more detail.

{\bf A)}
The term $\sum_i iz^\dag_i (\prt_t + ia_0 - i e A_0\si_3) z_i$
cannot appear in the effective Lagrangian. Note that our theory \Eq{o3Heff}
is invariant under a $90^\circ$ rotation followed by
$\v n_{ij} \to -\v n_{ij} $ (or $\v n_i \to -\v n_i$) transformation.
(Note $\Del^d_{ij}$ changes sign under the  $90^\circ$ rotation.)
Thus the effective theory \Eq{Leff} should be invariant under
a $90^\circ$ rotation followed by $z_i \to i\si_2 z_i^*$ transformation.
The linear time derivative term
$\sum_i iz^\dag_i \prt_t z_i$ changes sign under the above transformation and
cannot appear in the $L_{eff}$.

{\bf B)}
The effective Lagrangian \Eq{Leff} is invariant under the
$SU(2)$ spin rotation $W$:
\be
\pmatrix{\la_{1i} \cr (-)^i \la^\dag_{2i} \cr} \to W \pmatrix{\la_{1i} \cr
(-)^i \la^\dag_{2i} \cr}
\ee
moreover, when $A_\mu=0$, it is also invariant
under the $SU(2)$ pseudo-spin rotation:
\be
z_i \to U z_i
\ee
Under the pseudo-spin rotation, $A_\mu$ transform as
a component of a pseudo-spin vector. Thus
$A_\mu$ only couple to operators with pseudo-spin quantum number $(\tilde{S},
\tilde{S}_z)=(1,0)$. This result has an important consequence. The ground state
of the quantum disorder phase is a pseudo-spin singlet. Since $\psi$ is a
pseudo-spin singlet,  all the pseudo-spin 1 states have finite energy gaps.
The electromagnetic
field $A_\mu$ only connect the ground state to pseudo-spin 1 excited
states.  Thus the conductivity $\si(\om)$ is zero when $\om$ is less the
energy
gap for the pseudo-spin 1 excitations which is finite. Hence the quantum
disorder phase is an insulator with finite charge gap. The value of the charge
gap is given by $\Del_c$ in $L_{eff}$.
The argument should still hold even if some small pseudo-spin breaking
terms are introduced.

{\bf C)}
The $L_{eff}$ in \Eq{Leff} is very similar to the effective Lagrangian of
the $SU(2)$ slave-boson theory in the sF(staggered flux) phase. They both
contain spinons and holons. In the sF phase, the $SU(2)$ gauge field in the
$SU(2)$ slave-boson theory break down to a $U(1)$ gauge field which is just the
$U(1)$ gauge field in $L_{eff}$. Also, both the $SU(2)$ slave-boson theory and
the $L_{eff}$ contain two kinds of holons. It is very satisfying to see that
the $SU(2)$ slave-boson theory derived in the limit charge-gap $>$ spin-gap
smoothly connects to our effective theory $L_{eff}$ derived in the limit
charge-gap $<$ spin-gap.

We would like to stress that the above results can be valid even without the
exact $SU(2)$ pseudo-spin symmetry. For example the term
$\sum_{\vev{ij}} (n_i-1)(n_j-1) V_{ij}
= 4\sum_{\vev{ij}}\tilde{S}_{iz}\tilde{S}_{jz} V_{ij}$
breaks the $SU(2)$
pseudo-spin symmetry (Here $n_i$ is the particle number operator).
However, the term is invariant under the particle-hole
transformation which is realized by the 180$^\circ$ pseudo-spin rotation around
the $x$-axis. The gauge potential $A_\mu$ is odd and $\psi$ is even under such
a transformation.
Thus we can use the same symmetry arguments as done earlier, but this time
employing particle-hole symmetry instead of pseudo-spin symmetry, to show that
$\si(\om)=0$ for small $\om$ at half filling.  One can also show that
$\si(\om)=0$ even for non-zero chemical potential, as long as the ground state
remains to be the half filled state. Although it breaks the particle-hole
symmetry, the chemical potential term commutes with the Hamiltonian and cannot
change any matrix elements. Hence the chemical potential term cannot change
$\si(\om)$ as long as the ground state is not changed.

Now let us consider the situations with finite doping. First we
note that a finite hole density can be generated only when the
chemical potential is larger than the gap: $|eA_0|>\Del_c$.
Second, when $|eA_0|\sim \Del_c$, the $z_i$ field has a low
frequency solution which represent the low energy excitations. Let
us assume $eA_0 \sim \Del_c$ and replace
$eA_0$ by $\Del_c + eA_0$ (now $A_0$ is small). Expanding the
temporal term $|(\prt_t - ia_0 - i (\Del_c + e A_0) \si_3) z_i|^2 -
\Del_c^2 z^\dag_i z_i $, we get $ 2i\Del_c z^\dag_i \si_3 (\prt_t
-ia_0 -i eA_0\si_3 ) z_i $ after dropping the $ |(\prt_t -i
a_0)z|^2$ and $A_0^2$ term which are small as far as the low
energy excitations are concerned. Now the effective Lagrangian
\Eq{Leff} can be rewritten as
\ba
L_{eff} &=& \sum_i i\psi^\dag_i (\prt_t + ia_0 \si_3 ) \psi_i \nn\\
&&-\sum_{\vev{ij}} \left(  \psi^\dag_i (t_{ij}- i\Del^d_{ij} \si_3 (-)^i)
e^{ia_{ij}\si_3} \psi_j + h.c. \right) \nn\\ &+& \sum_i \left( \frac{i}{g'}
\Del_c z^\dag_i \si_3 (\prt_t - ia_0 - i e A_0\si_3) z_i \right)  \nn\\
&&-\frac{2}{g} \sum_{\vev{ij}} \left(  z^\dag_i
(e^{-ia_{ij}-i\frac{e}{c}A_{ij}\si_3}-\delta_{ij}) z_j + h.c. \right)  + ...
\label{Leffx}
\ea
which can be regarded as the effective theory for finite
doping. We see that $z$ now behaves as an ordinary (non-relativistic) boson
field. A finite density of $z$ can lead to a boson condensation which gives us
the $d$-wave superconducting state. The form and the physics of
\Eq{Leffx} are identical to that of the $SU(2)$ slave-boson theory at low
energies. Many results for the $SU(2)$ slave-boson theory \cite{WL,su2ph} also
apply to our effective theory here. In particular, our effective theory
(\Eq{Leff}, \Eq{Leffx}) explains the Mott insulating behavior at half
filling, as well as the spin gap, low superfluid density, and the positive
charge carriers above $T_c$ for underdoped samples. It is highly non-trivial to
recover the above properties within a single consistent theory.


XGW is supported by NSF Grant No. DMR--97--14198.  CLW, CYM and DC are
supported by grants from National Science Council of R.O.C (Taiwan).
We also like to thank NCTS for organizing the SCES topical program.


\end{document}